\documentclass[a4paper,12pt]{article}

\usepackage{graphicx}

\usepackage{amsmath}
\usepackage{amssymb}
\usepackage{cite}
\usepackage{color}
\usepackage{physics}
\usepackage{subcaption}

\allowdisplaybreaks

\newcommand{\mc}{\mathcal}

\newcommand{\del}{\partial}

\newcommand{\df}{\mathrm{d}}

\newcommand{\lag}{\mc{L}}
\newcommand{\ham}{\mc{H}}

\numberwithin{equation}{section}

\begin{document}

\renewcommand{\thefootnote}{\arabic{footnote}}
\newcommand{\bhline}[1]{\noalign{\hrule height #1}}
\newcommand{\bvline}[1]{\vrule width #1}

\setcounter{footnote}{0}

%%%%%%%%%%%%%%%%%%%%%%%%%%%

\renewcommand{\thefootnote}{\fnsymbol{footnote}}

\begin{flushright}
KUNS-2864
\end{flushright}
\vspace*{0.5cm}

\begin{center}
{\Large \bf Chaotic string dynamics in deformed $T^{1,1}$   
}
\vspace*{2cm} \\
{\large  Takaaki Ishii\footnote{E-mail:~ishiitk@gauge.scphys.kyoto-u.ac.jp},
Shodai Kushiro\footnote{E-mail:~kushiro@gauge.scphys.kyoto-u.ac.jp},
and Kentaroh Yoshida\footnote{E-mail:~kyoshida@gauge.scphys.kyoto-u.ac.jp}} 
\end{center}

\vspace*{0.4cm}

\begin{center}
{\it Department of Physics, Kyoto University, Kyoto 606-8502, Japan.}
\end{center}

\vspace{2cm}

\begin{abstract}
Recently, Arutyunov, Bassi and Lacroix have shown that 2D non-linear sigma model with a deformed $T^{1,1}$ background is classically integrable [arXiv:2010.05573 [hep-th]]. This background includes a Kalb-Ramond 
two-form with a critical value. Then the sigma model has been conjectured to be non-integrable 
when the two-form is off critical. We confirm this conjecure by explicitly presenting classical chaos.  
With a winding string ansatz, the system is reduced to a dynamical system described by a set of ordinary differential equations. 
Then we find classical chaos, which indicates non-integrability, by numerically computing Poincar\'{e} sections and Lyapunov spectra 
for some initial conditions.
\end{abstract}

\setcounter{footnote}{0}
\setcounter{page}{0}
\thispagestyle{empty}

\newpage

\tableofcontents

\renewcommand\thefootnote{\arabic{footnote}}

%%%%%%%%%%%%%%%%
\section{Introduction}
\label{sec:intro} 

The AdS/CFT correspondence \cite{Maldacena:1997re,Witten:1998qj,Gubser:1998bc} is one of the most significant topics 
in String Theory. A typical example is the duality between type IIB string theory on 
AdS$_{5}\times$S$^{5}$ and the $\mc{N}=4$ $SU(N)$ super Yang-Mills theory in large $N$ limit. 
A possible generalization, preserving the conformal symmetry of the boundary theory, is to replace the internal manifold S$^5$ by an Einstein manifold 
$T^{1,1}$ \cite{Page:1984ae,Romans:1984an,Candelas:1989js}. 
The metric of $T^{1,1}$ is given by 
\begin{equation}
\df s^2 = 
\frac{1}{6} \sum^{2}_{i=1}
 \left( \df \theta_i^2+\sin^2 \theta_i\df \phi_i^2\right)
 + \frac{1}{9} \left(\df \psi +\cos \theta_1 \df \phi_1+\cos \theta_2 \df \phi_2\right)^2
\label{T11}
\end{equation} 
(in units that the dimension of the curvature scale is unity). 
The AdS$_5\times T^{1,1}$ geometry has been elaborated in \cite{Klebanov:1998hh}.
In particular, the classical dynamics of a string moving on $T^{1,1}$ is chaotic and hence non-integrable \cite{Basu:2011di}. 
This non-integrability has been shown also in an analytic manner \cite{Basu:2011fw}. Furthermore, the chaotic dynamics 
has been shown in a near Penrose limit \cite{Asano:2015qwa}. The coset construction of 
the $T^{1,1}$ metric based on the coset 
\begin{equation}
\frac{SU(2)_{\rm L}\times SU(2)_{\rm R}\times U(1)}{U(1)_{\rm L}\times U(1)_{\rm R}} 
\end{equation}
has been discussed in \cite{Crichigno:2014ipa}. Yang-Baxter deformations \cite{Klimcik1,Klimcik2, DMV1,DMV2,KMY1,MY} of $T^{1,1}$ are discussed in \cite{Crichigno:2014ipa,SY-T11,Rado:2020yhf} 
(For a short summary, see \cite{Crichigno:2015sga}).

\medskip 

Recently, an intriguing generalization of the $T^{1,1}$ background has been considered by Arutyunov, Bassi and Lacroix 
\cite{Arutyunov:2020sdo}. The background is given by
\begin{subequations} 
 \label{eq:T11B}
\begin{align}
\df s^2
 &=\sum^{2}_{i=1}\lambda^{2}_{i}
 \left( \df \theta_i^2+\sin^2 \theta_i\df \phi_i^2\right)
 +\lambda^2\left(\df \psi +\cos \theta_1 \df \phi_1+\cos \theta_2 \df \phi_2\right)^2\,, 
 \label{eq:metric}\\
 B_2&
 =k\left(\df \psi+\cos \theta_1 \df \phi_1\right)\wedge\left(\df \psi+\cos \theta_2 \df \phi_2\right)
 \label{eq:bfield}
\end{align}
\end{subequations}
and contains four real parameters $\lambda_{1}$\,, $\lambda_{2}$\,, $\lambda$ and $k$\,. 
We shall denote this background by the symbol $\mathcal{T}_{k}^{(\lambda_1,\lambda_2,\lambda)}$\,. 
This metric (\ref{eq:metric}) is not Einstein in general. 
Notice that this background contains a Kalb-Ramond two-form $B_2$\,. 
\medskip 

It should be remarked that when $k=\lambda^2$\,, some interesting things happen.  
The classical string dynamics on $R\times \mathcal{T}_{\lambda^2}^{(\lambda_1,\lambda_2,\lambda)}$\,,
where $R$ describes the time direction in target space, is integrable \cite{Arutyunov:2020sdo} 
in the sense of Lax pair. 
The special case with $\lambda_{1}^{2}=\lambda_{2}^{2}=\frac{3}{2}\lambda^{2}$ 
gives rise to the original $T^{1,1}$ metric (\ref{T11})\,. Note that even for this original metric, 
the integrability is still preserved due to the existence of the $B$-field (\ref{eq:bfield})\,. 
Also, the case with $\lambda_{1}=\lambda_{2}=\lambda$ has already been known 
as the Guadagnini-Martellini-Mintchev (GMM) model \cite{Guadagnini:1987ty} 
and used to construct an NS-NS supergravity background with the flux $k=\lambda^2$ \cite{PandoZayas:2000he}. 
For recent progress on the RG flow in this model, 
see \cite{Levine:2021fof}.

\medskip 

On the other hand, it is conjectured that the system would be non-integrable 
when $k\neq \lambda^2$ \cite{Arutyunov:2020sdo}\footnote{In this sense, the particular value $k=\lambda^2$ may be called critical and otherwise off critical.}. Hence it should be significant 
to show this non-integrability by following some standard manners.   
In this letter, we will consider cases with $k\neq \lambda^2$ and 
show that the string sigma model on $R\times \mathcal{T}_{k}^{(\lambda_1,\lambda_2,\lambda)}$ exhibits classical chaos in string motion.
To reduce the sigma model to a dynamical system, we employ a winding string ansatz 
in which a string is wrapped around all of the isometry directions, i.e.~$\phi_1, \, \phi_2$ and $\psi$\,. 
Then the chaotic behaviors are explicitly shown 
by numerically computing Poincar\'{e} sections and Lyapunov spectra for some initial values. 

\medskip 

This letter is organized as follows. In section 2, we introduce the classical string sigma model action 
on $R\times\mathcal{T}_{k}^{(\lambda_1,\lambda_2,\lambda)}$\,. Then the system is reduced by using a winding string ansatz. 
In section 3, Poincar\'{e} sections and Lyapunov spectra are computed numerically for some initial values and classical chaos appears when $k\neq \lambda^2$\,. 
Section 4 is devoted to conclusion and discussion.

%%%%%%%%%%%%%%%%
\section{Reducing string sigma model on 
$R\times\mathcal{T}_{k}^{(\lambda_1,\lambda_2,\lambda)}$}
\label{sec:2}
%%%%%%%%%%%%%%%%

We shall consider a classical string on the background 
$R\times\mathcal{T}_{k}^{(\lambda_1,\lambda_2,\lambda)}$\,. 
As discussed in \cite{Arutyunov:2020sdo}, the original isometries as $T^{1,1}$ are retained in the presence of $B$-field.
In particular, one can easily see in \eqref{eq:T11B} that constant shifts in $(\phi_1,\phi_2,\psi)$-directions are still symmetries.

\medskip 

Note that we suppose that the background $R\times\mathcal{T}_{k}^{(\lambda_1,\lambda_2,\lambda)}$ should be obtained from a supergravity embedding of the background (\ref{eq:T11B})\,, though it has not been done other than 
for special values of the parameters. So this is just analogy of $R\times S^3$ in the case of AdS$_5\times$S$^5$\,, 
where $R$ comes from AdS time.  

%%%%%%%%%%
\subsection{String sigma model on $R\times\mathcal{T}_{k}^{(\lambda_1,\lambda_2,\lambda)}$}
\label{sec:2.1}
%%%%%%%%%%

Let $G_{\mu\nu}$ and $B_{\mu\nu}$ $(\mu,\nu=t,\theta_{1},\phi_{1},\theta_{2},\phi_{2},\psi)$ denote the background metric and $B$-field, $h_{ab}$ $(a,b=\tau,\sigma)$ the world-sheet metric, and $X^{\mu}$ the target-space coordinates. 
The classical action of the string $\sigma$-model is given by
\begin{equation}
 S =  -\frac{1}{2} \int \df \tau \df \sigma
 \left( \sqrt{-h} h^{ab} G_{\mu\nu}+ \epsilon^{ab} B_{\mu\nu} \right) \del_{a}X^{\mu}\del_{b}X^{\nu}\,,
\label{eq:action}
\end{equation}
where $h \equiv \mathrm{det} \, h_{ab}$\, and the antisymmetric symbol is defined as $\epsilon^{\tau\sigma}=-1$\,.
The equation of motion for $X^\mu$ is given by
\begin{equation}
\partial_a \partial^a X^\mu + \Gamma^\mu_{\rho\sigma} h^{ab} \partial_a X^\rho \partial_b X^\sigma - \mathcal{T}{}^\mu_{\rho\sigma} \varepsilon^{ab} \partial_a X^\rho \partial_b X^\sigma=0\,,
\label{eq:eomX}
\end{equation}
where $\varepsilon^{ab} \equiv \epsilon^{ab}/\sqrt{-h}$ and
\begin{equation}
\Gamma^\mu_{\rho\sigma} = \frac{1}{2} G^{\mu\nu}\left( G_{\nu\sigma,\rho} + G_{\rho\nu,\sigma} - G_{\rho\sigma,\nu} \right)\,, \quad
\mathcal{T}{}^\mu_{\rho\sigma} = \frac{1}{2} G^{\mu\nu}\left( G_{\rho\sigma,\nu} + G_{\nu\rho,\sigma} + G_{\sigma\nu,\rho} \right)\,.
\end{equation}
The equation of motion for $h^{ab}$ imposes that the stress-energy tensor $T_{ab}$ on the world-sheet should vanish: 
\begin{equation}
0 = T_{ab}=G_{\mu\nu}\partial_a X^{\mu} \partial_b X^{\nu}  -\frac{1}{2} h_{ab} h^{cd} G_{\mu\nu}\partial_c X^{\mu} \partial_d X^{\nu}\,.
\label{eq:Tab}
\end{equation}
Note here that  the variation of $h^{ab}$ is irrelevant to the antisymmetric symbol. 
The vanishing stress-energy tensor  leads to the Virasoro constraints which restrict the string embedding 
specified by $X^{\mu}$\,.

\medskip

Thanks to the world-sheet diffeomorphism, one can take the conformal gauge $h_{ab} = e^{2\omega(\tau,\sigma)} \eta_{ab}$\,, where $\eta_{ab}=\mathrm{diag}(-1,1)$\,.
In this gauge, the action takes the form: 
\begin{equation}
 S^{(\rm{c})} =\int \df \tau \df \sigma \, \lag = -\frac{1}{2} \int \df \tau \df \sigma
 \left( \eta^{ab} G_{\mu\nu}+ \epsilon^{ab} B_{\mu\nu} \right) \del_{a}X^{\mu}\del_{b}X^{\nu}\,.
\label{eq:action2}
\end{equation}
The conjugate momentum is given by 
\begin{align}
p_{\mu} 
=\frac{\del \lag}{\del \left(\del_{\tau} X^{\mu} \right)}=G_{\mu\nu}\del_{\tau}X^{\nu}+B_{\mu\nu}\del_{\sigma}X^\nu\,.
\label{eq:pmu}
\end{align}
Then the Hamiltonian is given by  
\begin{align}
\ham&=p_{\mu}\partial_\tau X^{\mu}-\lag
=\frac{1}{2}G_{\mu\nu}\del_{\tau} X^{\mu} \del_{\tau} X^{\nu}-\frac{1}{2}G_{\mu\nu}\del_{\sigma} X^{\mu} \del_{\sigma} X^{\nu}\,.
\end{align}
While the $B$-field appears in \eqref{eq:pmu}, it cancels with the $B$-field from $\lag$ in the above expression.
As a result, the canonical stress-energy tensor made of the conjugate momentum gives the same result as \eqref{eq:Tab} as it can be checked easily.

\medskip

Once the gauge has been fixed, the Virasoro constraints can be expressed as follows.
The $(\tau,\tau)$-component of the stress-energy tensor is precisely the Hamiltonian,
i.e. $T_{\tau\tau}=\ham$ and hence 
the Virasoro constraints indicate that $\ham=0$ (the Hamiltonian constraint)\,.
We also find that $T_{\tau\tau}=T_{\sigma\sigma}$ because $T^a{}_a=0$\,.  
Substituting \eqref{eq:pmu} into \eqref{eq:Tab} in the conformal gauge, the other independent component $T_{\tau\sigma}$ 
can be rewritten as 
\begin{align}
T_{\tau\sigma}=p_{\mu}\partial_\sigma X^{\mu}=0\,.
\label{eq:tsVirasoro}
\end{align}

\medskip

When $k=\lambda^2$\,, this string sigma model has a classical Lax pair and is classically integrable \cite{Arutyunov:2020sdo}. 
Our purpose here is to see the anticipated non-integrability when $k\neq\lambda^2$\,.

%%%%%%%%%%%%%%%%%%%%%
\subsection{Reduction of the system with a winding string ansatz}
\label{sec:2.2}
%%%%%%%%%%%%%%%%%%%%%

In order to show non-integrability, it is sufficient to find out chaotic motions in a subsector. 
Hence it is convenient to consider the simplest subsector which contains chaotic motions. 
To reduce the system to a set of ordinary differential equations, 
let us suppose the following winding string ansatz:  
\begin{equation}
\begin{split}
t&=t(\tau)\,,\quad
\theta_{1}=\theta_{1}(\tau)\,,\quad
\theta_{2}=\theta_{2}(\tau)\,,\\
\phi_{1}&=\phi_{1}(\tau)+\alpha_{1}\sigma\,,\quad
\phi_{2}=\phi_{2}(\tau)+\alpha_{2}\sigma\,,\quad
\psi=\psi(\tau)+\alpha \sigma\,.
\end{split}
\label{eq:ansatz}
\end{equation}
This ansatz describes a string wrapped on $(\phi_1,\phi_2,\psi)$-directions.
Since the periodicity of the compact coordinates are $\phi_{i} \simeq \phi_{i} + 2 \pi$ $(i=1,2)$ and $\psi \simeq \psi+4 \pi$\,, 
and $\sigma\simeq\sigma+2\pi$\,,  
the winding numbers $\alpha_i~(i=1,2)$ and $\alpha$ should satisfy $\alpha_{i}, \alpha/2 \in \mathbb{Z}$\,.

\medskip

By substituting the ansatz \eqref{eq:ansatz} into \eqref{eq:pmu},
the momentum components associated with the isometries (and hence conserved) are given by 
\begin{subequations}
\label{eq:p_isometries}
\begin{align}
p_t &= -\dot{t} \,,\\
p_{\phi_{1}}&=
\qty(\lambda_{1}^{2}\sin^2 \theta_{1} +\lambda^{2}\cos^{2} \theta_{1})\dot{\phi_{1}} \notag \\ 
&\quad +\lambda^{2}\cos\theta_{1}\cos\theta_{2}\dot{\phi_{2}}
+\lambda^{2}\cos\theta_{1}\dot{\psi}
+k \cos \theta_{1} \qty(\alpha_{2}\cos\theta_{2}+\alpha)\,,\\
p_{\phi_{2}}&=
\qty(\lambda_{2}^{2}\sin^2 \theta_{2} +\lambda^{2}\cos^{2} \theta_{2})\dot{\phi_{2}} \notag \\ 
& \quad +\lambda^{2}\cos\theta_{1}\cos\theta_{2}\dot{\phi_{1}}
+\lambda^{2}\cos\theta_{2}\dot{\psi}
-k \cos \theta_{2} \qty(\alpha_{1}\cos\theta_{1}+\alpha)\,,\\
p_{\psi}&=
\lambda^2 \cos\theta_{1}\dot{\phi_{1}}
+\lambda^2 \cos\theta_{2}\dot{\phi_{2}}
+\lambda^2\dot{\psi}
-k\alpha_{1}\cos\theta_{1}
+k\alpha_{2}\cos\theta_{2}\,.
\end{align}
\end{subequations} 
Under the ansatz \eqref{eq:ansatz}\,,  the constraint \eqref{eq:tsVirasoro} can be rewritten as 
\begin{equation}
T_{\tau\sigma}=p_{\phi_{1}}\alpha_{1}+p_{\phi_{2}}\alpha_{2}+p_{\psi}\alpha=0\,.
\end{equation}
From this expression, one finds that this constraint can be satisfied by simply taking $p_{\phi_{1}}=p_{\phi_{2}}=p_{\psi}=0$\,. 
It is also helpful to  introduce the energy $E$ defined as $E \equiv -p_t$\,.
Then, from \eqref{eq:p_isometries}, we obtain
\begin{subequations}
\label{eq:isometry}
\begin{align}
\dot{t}&=E\,,\\
\dot{\phi_{1}}&=-\frac{k\cos \theta_{1}\left(\alpha+\alpha_{1}\cos\theta_{1}\right)}{\lambda_{1}^{2}\sin^2\theta_{1}}\,,\\
\dot{\phi_{2}}&=\frac{k\cos \theta_{2}\left(\alpha+\alpha_{2}\cos\theta_{2}\right)}{\lambda_{2}^{2}\sin^2\theta_{2}}\,,\\
\dot{\psi}&=k\left(\frac{\alpha_{1}\cos\theta_{1}-\alpha_{2}\cos\theta_{2}}{\lambda^2}+\frac{\cos^{2} \theta_{1} \left(\alpha+\alpha_{1} \cos\theta_{1} \right)}{\lambda^{2}_{1}\sin^{2} \theta_{1}}-\frac{\cos^{2} \theta_{2} \left(\alpha+\alpha_{2} \cos\theta_{2} \right)}{\lambda^{2}_{2}\sin^{2} \theta_{2}}\right)\,.
\end{align}
\end{subequations} 
The other components are given by
\begin{equation}
p_{\theta_i}= \lambda_i^2 \dot{\theta}_i\,.
\end{equation}

\medskip

The string motion is described by coupled ordinary differential equations for $\theta_{i}$.
Substituting \eqref{eq:ansatz} into \eqref{eq:eomX} and using \eqref{eq:isometry}, we obtain the  evolution equations for $\theta_{i}$ as 
\begin{subequations}
\label{system}
\begin{align}
\frac{\ddot{\theta_{1}}}{\sin\theta_{1}}=\alpha_{1}^2\left(\frac{\lambda^2}{\lambda^{2}_{1}}-1+\frac{k^2}{\lambda^2\lambda_{1}^{2}}\left(1-\frac{\lambda^2}{\lambda^{2}_{1}}+\frac{\lambda^2}{\lambda^{2}_{1}\sin^{4}\theta_{1}}\right)\right)\cos\theta_{1}-\alpha_{1}\alpha_{2}\frac{k^2-\lambda^4}{\lambda^2\lambda_{1}^{2}}\cos\theta_{2} \notag\\
+\frac{\alpha\alpha_{1}}{\lambda_{1}^{4}}\left(\lambda^2\lambda^{2}_{1}-k^{2}+\frac{k^{2}\left(\cos^2\theta_{1}+1\right)}{\sin^4\theta_1}\right)+\frac{\alpha^{2}k^{2}\cos\theta_{1}}{\lambda^{4}_{1}\sin^{4}\theta_{1}}\,,\\
\frac{\ddot{\theta_{2}}}{\sin\theta_{2}}=\alpha_{2}^2\left(\frac{\lambda^2}{\lambda^{2}_{2}}-1+\frac{k^2}{\lambda^2\lambda_{2}^{2}}\left(1-\frac{\lambda^2}{\lambda^{2}_{2}}+\frac{\lambda^2}{\lambda^{2}_{2}\sin^{4}\theta_{2}}\right)\right)\cos\theta_{2}-\alpha_{1}\alpha_{2}\frac{k^2-\lambda^4}{\lambda^2\lambda_{2}^{2}}\cos\theta_{1} \notag\\
+\frac{\alpha\alpha_{2}}{\lambda_{2}^{4}}\left(\lambda^2\lambda^{2}_{2}-k^{2}+\frac{k^{2}\left(\cos^2\theta_{2}+1\right)}{\sin^4\theta_2}\right)+\frac{\alpha^{2}k^{2}\cos\theta_{2}}{\lambda^{4}_{2}\sin^{4}\theta_{2}}\,.
\end{align}
\end{subequations}
Note that the two equations are decoupled when $k=\lambda^2$ 
and no chaos appears obviously.  

\medskip

The Hamiltonian becomes 
\begin{align}
\ham 
=-\frac{1}{2}E^2+\frac{1}{2\lambda_{1}^{2}}p_{\theta_{1}}^2+\frac{1}{2\lambda_{2}^{2}}p_{\theta_{2}}^2 + V\qty(\theta_{1},\theta_{2})\,,
\end{align}
where the potential $V(\theta_1,\theta_2)$ is defined as 
\begin{align}
V\qty(\theta_{1},\theta_{2}) \equiv\,&\frac{\alpha_{1}^2}{2\lambda^2\lambda_{1}^2}\frac{\left(\lambda^2\cos^2\theta_1+\lambda_{1}^{2}\sin^2\theta_{1}\right)\left(k^2\cos^2\theta_1+\lambda^2\lambda_{1}^{2}\sin^2\theta_{1}\right)}{\sin^2\theta_{1}} \nonumber\\
&+\frac{\alpha_{2}^2}{2\lambda^2\lambda_{2}^2}\frac{\left(\lambda^2\cos^2\theta_2+\lambda_{2}^{2}\sin^2\theta_{2}\right)\left(k^2\cos^2\theta_2+\lambda^2\lambda_{2}^{2}\sin^2\theta_{2}\right)}{\sin^2\theta_{2}} \nonumber\\
&-\alpha_{1}\alpha_{2}\frac{k^2-\lambda^4}{\lambda^2}\cos\theta_1\cos\theta_2\nonumber\\
&+\frac{\alpha^2}{2}\left(\lambda^2+\frac{k^2 \cot^2\theta_{1}}{\lambda_{1}^{2}}+\frac{k^2 \cot^2\theta_{2}}{\lambda_{2}^{2}}\right)\nonumber\\
&+\frac{\alpha\alpha_{1}}{\lambda_{1}^{2}}\qty(k^2\cot^2\theta_{1}+\lambda^2\lambda^{2}_{1})\cos\theta_{1}+\frac{\alpha\alpha_{2}}{\lambda_{2}^{2}}\qty(k^2\cot^2\theta_{2}+\lambda^2\lambda^{2}_{2})\cos\theta_{2}\,. 
\end{align}
Recall that the Hamiltonian has to be zero, i.e. $\ham=0$, due to the Virasoro constraints.
Since the Hamiltonian is conserved, 
this constraint is always satisfied along time evolution 
once the initial data is prepared so as to satisfy $\ham=0$\,. 

\medskip

The presence of the background $B$-field restricts the domain of the string motion.
When $k \neq 0$\,, the string cannot reach the poles of $S^2$ ($\theta_{i}=0$ and $\pi$), 
because $V\qty(\theta_{1},\theta_{2})$ diverges there.
Therefore, the string can move in the range $0<\theta_{i}<\pi$\,.
When $k=0$\,, the divergent terms are absent in the potential, 
and the string can shrink and pass through the poles. 

\medskip

Note also that the potential is tilted because of the winding in the $\psi$-direction 
(i.e., $\alpha\neq 0$)\,. 
If $\alpha=0$, the potential is symmetric under $(\theta_{1},\theta_{2}) \to (\pi-\theta_{1},\pi-\theta_{2})$, while it is not otherwise.
Hence, there is one global minimum when $\alpha\neq0$\,.
In practice, the minimum can be found numerically with specified parameters.

%%%%%%%%%%%%%%%%%%
\section{Chaotic string dynamics}
\label{sec:3}
%%%%%%%%%%%%%%%%%%

In this section, we present classical chaos in the reduced system (\ref{system}) 
by computing Poincar\'{e} sections and Lyapunov spectra for some values when $k \neq \lambda^2$\,.  

\medskip 

For numerical computations, we set the parameters as 
\begin{eqnarray}
\alpha_1=\alpha_2=1\,, \quad  \alpha=2\,, \quad \lambda_1=\lambda_2=\lambda=1
\end{eqnarray}
for simplicity. 
We have examined various values of $k$, but here we will show results for the cases that 
$k=0$\,, $0.5$ and $10$\,. 
For other values of $k$ as far as we have examined, classical chaos has been observed 
(except the special case with $k=\lambda^2$)\,.
Poincar\'{e} sections are computed for 
\begin{eqnarray}
\theta_2=\pi~(k=0) \quad \mbox{or} \quad \theta_2=\pi/2~(k=0.5,\,10) \quad \mbox{and} \quad p_{\theta_2}\geq0\,. 
\end{eqnarray}
In each figure, the energy $E$ is fixed and the initial conditions for the differential equations are changed. 
Different colors correspond to different initial conditions.

\medskip

In addition, to measure the strength of chaos, we have examined the Lyapunov spectrum as well 
(For the detail of the computation, for example, see \cite{Shimada:1979})\,.
In the system (\ref{system})\,, the Lyapunov spectrum contains four exponents and the largest one is 
the most significant for the growth of chaos.

\medskip

As a remark on $\alpha$\,, for $\alpha=0$ and $\lambda_1=\lambda_2=\lambda$, we could not find any chaos in string motion even for $k \neq \lambda^2$. This would be simply 
because this case corresponds to an integrable subsector. 

\medskip 

In the following, let us present Poincar\'{e} sections and Lyapunov spectra for $k=0$\,, $0.5$ and $10$\,, respectively. 

%%%%
\subsubsection*{(i) $k=0$ case}

\begin{figure}[htbp]
\centering
\begin{tabular}{c}
\begin{minipage}{0.45\hsize}
\centering
\includegraphics[width=7cm]{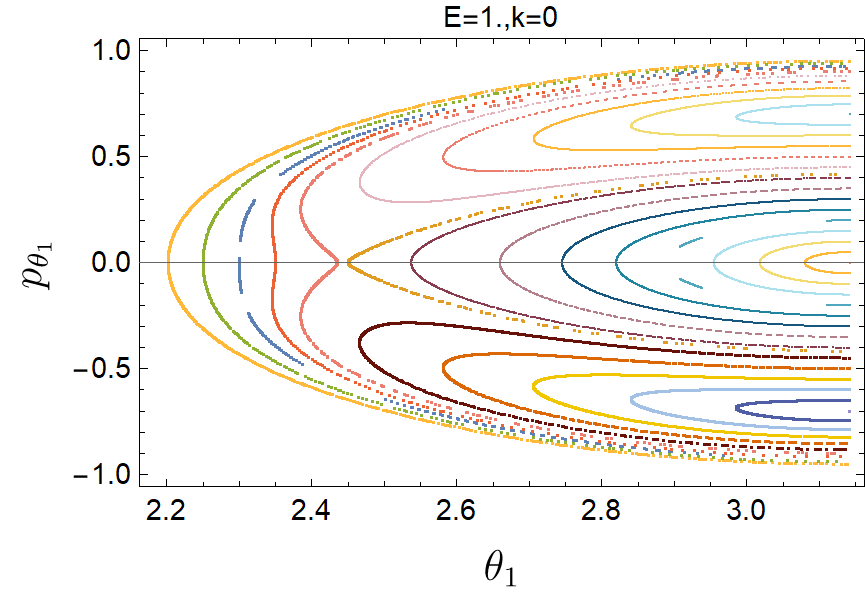}
\subcaption{\footnotesize Poincar\'{e} sections for $E=1$}
\label{fig:k0e1}
\end{minipage}
\begin{minipage}{0.45\hsize}
\centering
\includegraphics[width=7cm]{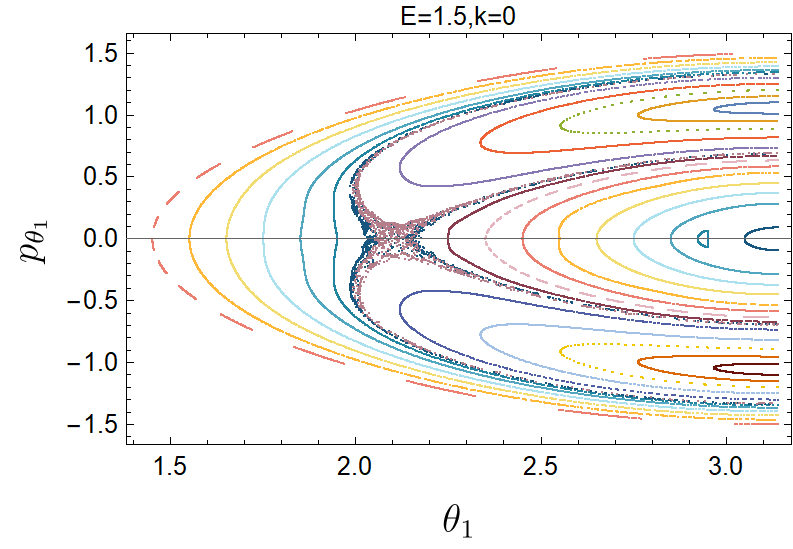}
\subcaption{\footnotesize Poincar\'{e} sections for $E=1.5$}
\label{fig:k0e1.5}
\end{minipage}
\vspace*{0.7cm}\\
\begin{minipage}{0.45\hsize}
\centering
\includegraphics[width=7cm]{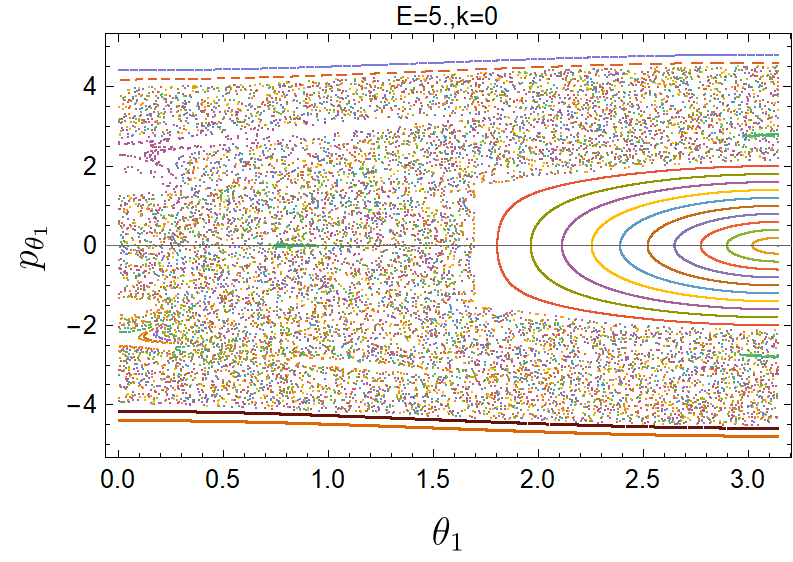}
\subcaption{\footnotesize Poincar\'{e} sections for $E=5$}
\label{fig:k0e5}
\end{minipage}
\begin{minipage}{0.45\hsize}
\centering
\includegraphics[width=7cm]{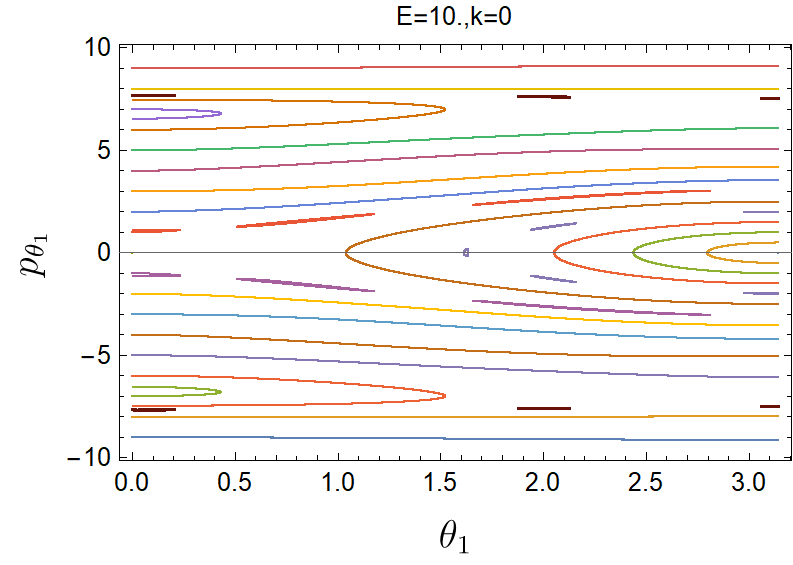}
\subcaption{\footnotesize Poincar\'{e} sections for $E=10$}
\label{fig:k0e10}
\end{minipage}
\vspace*{0.7cm}\\
\begin{minipage}{0.45\hsize}
\centering
\includegraphics[width=7cm]{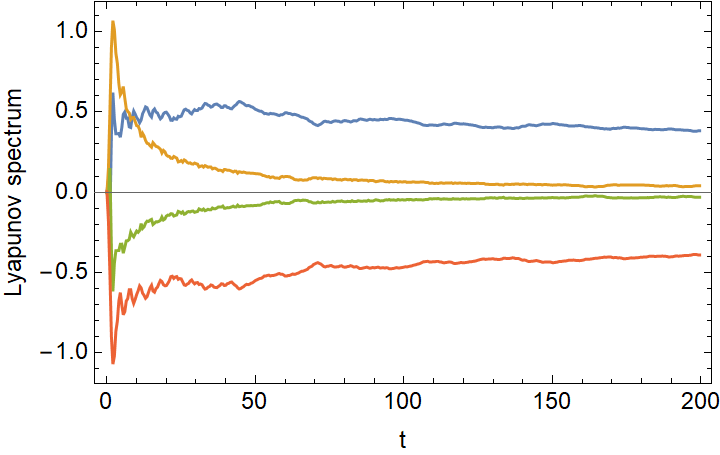}
\subcaption{\footnotesize Lyapunov spectrum for $E=5$, $p_{\theta_1}(0)=1$}
\label{fig:lyapuk0}
\end{minipage}
\end{tabular}
\vspace*{0.7cm}
\caption{Poincar\'{e} sections and Lyapunov spectrum for $k=0$}
\label{fig:k=0}
\end{figure}

Figure \ref{fig:k=0} shows Poincar\'{e} sections for $k=0$\,. 
The potential minimum is $(\theta_{1},\theta_{2})=(\pi,\pi)$\,. 
In the low-energy region $(E=1)$, there are only Kolmogorov-Arnold-Moser (KAM) tori \cite{Kolmogorov:1954,Arnold:1963,Moser:1962} 
and is no chaos [Fig.\,\ref{fig:k=0} (a)]\,. One can see a separatrix structure around $\theta_1=2.44$ and $p_{\theta_1}=0$\,. 
When $E=1.5$\,, the separatrix begins to collapse 
and classical chaos appears [Fig.\,\ref{fig:k=0} (b)]\,.
When $E=5$\,, a lot of KAM tori are broken and an island of KAM tori survives on the right-hand side [Fig.\,\ref{fig:k=0} (c)]\,. 
When $E=10$\,, all of the trajectories become KAM tori again and no chaos is found [Fig.\,\ref{fig:k=0} (d)]\,. 
This is a typical behavior of classical chaos in the high-energy region. 

\medskip

Figure \ref{fig:k=0} (e) is a Lyapunov spectrum for $E=5$\,. 
The initial values of $\theta_{1}(t)$, $\theta_{2}(t)$, $p_{\theta_{1}}(t)$ and $p_{\theta_{1}}(t)$ 
are taken as $\theta_{1}(0)=1$, $\theta_{2}(0)=\pi$, $p_{\theta_{1}}(0)=1$, and $p_{\theta_{2}}(0)\sim4.574$\,. 
The maximum Lyapunov exponent is evaluated as the average for $150 \le t \le 200$\,. 
Then it is $0.3990\pm0.0116$, where the errors are given by the standard deviation.

%%%%%%%%%%%%%%%%%%%%%

\subsubsection*{(ii) $k=0.5$ case}

\begin{figure}[htbp]
\centering
\begin{tabular}{c}
\begin{minipage}{0.45\hsize}
\centering
\includegraphics[width=7cm]{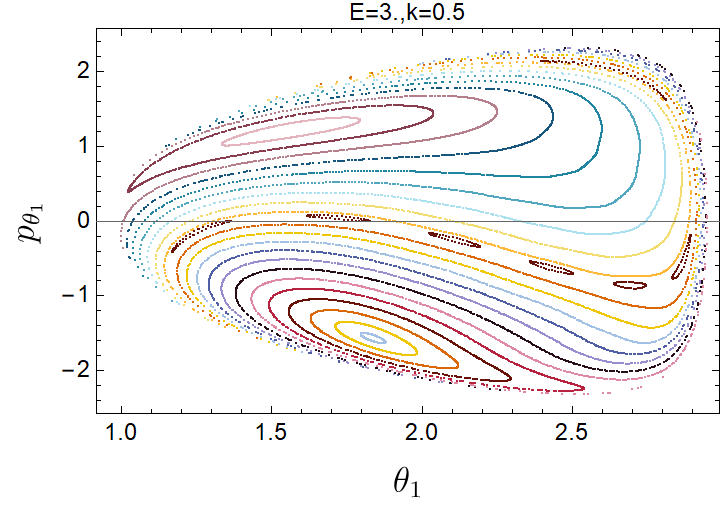}
\subcaption{\footnotesize Poincar\'{e} sections for $E=3$}
\label{fig:k0.5e3}
\end{minipage}
\begin{minipage}{0.45\hsize}
\centering
\includegraphics[width=7cm]{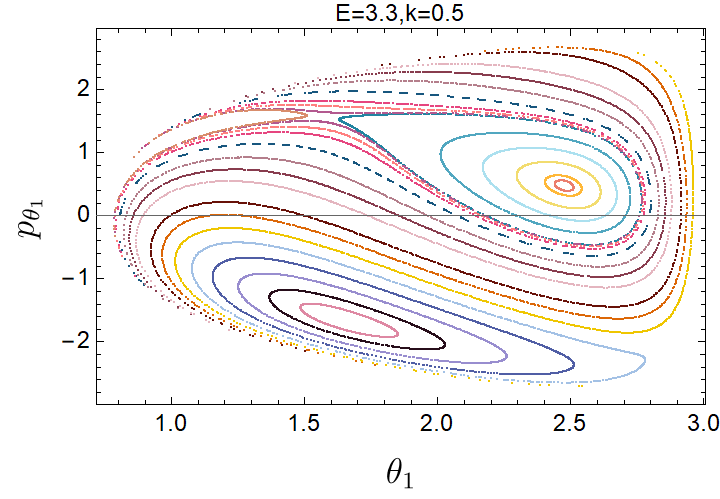}
\subcaption{\footnotesize Poincar\'{e} sections for $E=3.3$}
\label{fig:k0.5e33}
\end{minipage}
\vspace*{0.7cm}\\
\begin{minipage}{0.45\hsize}
\centering
\includegraphics[width=7cm]{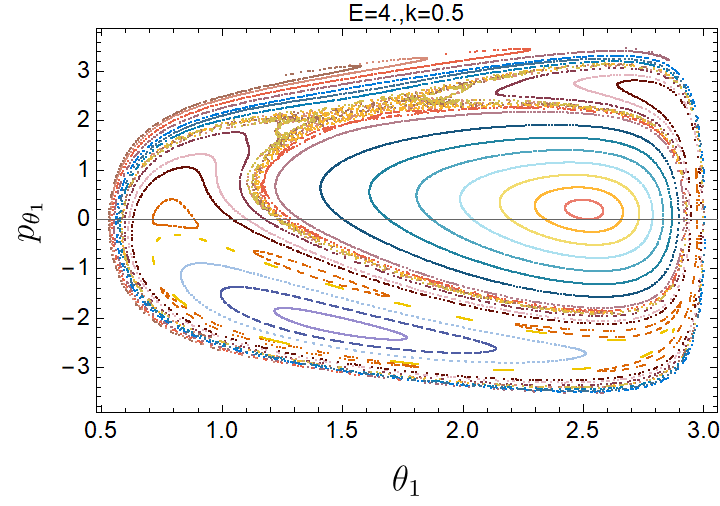}
\subcaption{\footnotesize Poincar\'{e} sections for $E=4$}
\label{fig:k0.5e4}
\end{minipage}
\begin{minipage}{0.45\hsize}
\centering
\includegraphics[width=7cm]{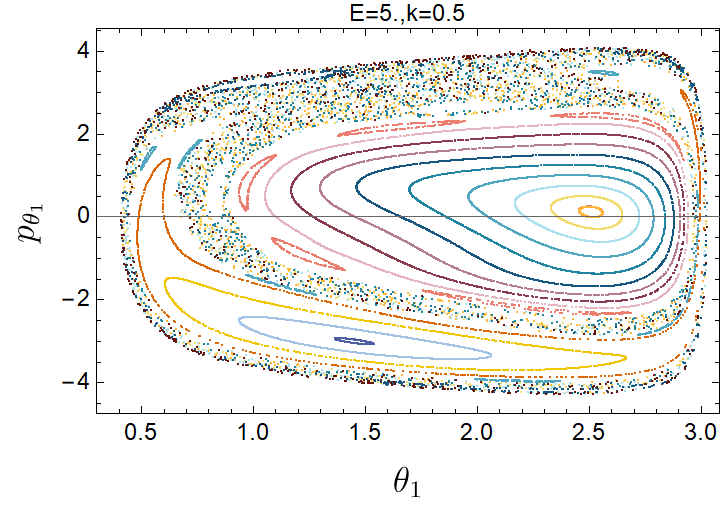}
\subcaption{\footnotesize Poincar\'{e} sections for $E=5$}
\label{fig:k0.5e5}
\end{minipage}
\vspace*{0.7cm}\\
\begin{minipage}[b]{0.45\hsize}
\centering
\includegraphics[width=7cm]{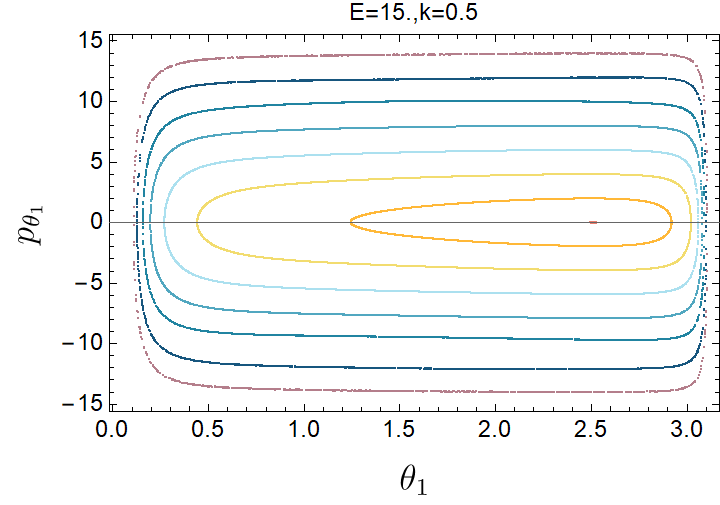}
\subcaption{\footnotesize Poincar\'{e} sections for $E=15$}
\label{fig:k0.5e15}
\end{minipage}
\begin{minipage}[b]{0.45\hsize}
\centering
\includegraphics[width=7cm]{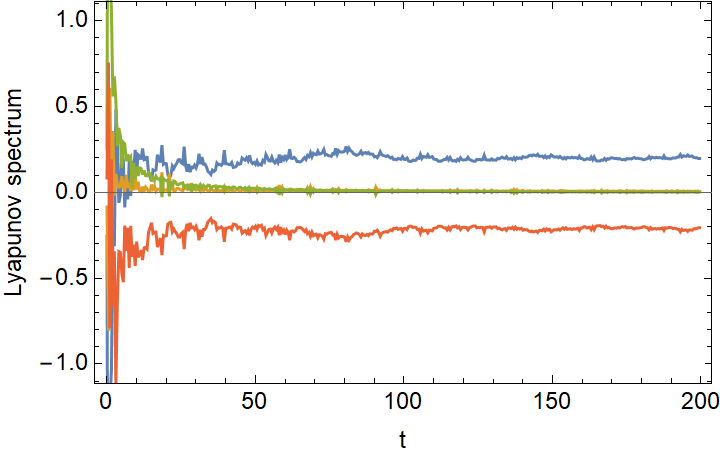}
\subcaption{\footnotesize Lyapunov spectrum for $E=5$, $p_1=3$}
\label{fig:lyapuk0.5}
\end{minipage}
\end{tabular}
\vspace*{0.7cm}
\caption{Poincar\'{e} sections and Lyapunov spectrum for $k=0.5$}
\label{fig:k=0.5}
\end{figure}

Figure \ref{fig:k=0.5} shows Poincar\'{e} sections for $k=0.5$\,.  
The potential minimum is $(\theta_{1},\theta_{2})\sim(2.438,2.438)$\,.
In the low-energy region $(E=3)$\,, there are only KAM tori only and is no chaos [Fig.\,\ref{fig:k=0.5} (a)]\,. 
As the energy is a bit increased to $E=3.3$\,, a separatrix structure appears around  $\theta_1=1.51$ and $p_{\theta_1}=1.6$ 
as shown in Fig.\,\ref{fig:k=0.5} (b)\,.  
When $E=4$\,, one can see that the separatrix begins to collapse and classical chaos appears [Fig.\,\ref{fig:k=0.5} (c)]\,.
When $E=5$\,, chaos is clearly observed while some islands of KAM tori survive[Fig.\,\ref{fig:k=0.5} (d)]\,.
In the higher energy region $(E=15)$\,, there are only KAM tori and the chaos disappears again [Fig.\,\ref{fig:k=0.5} (e)]\,.

\medskip

Figure \ref{fig:k=0.5} (f) is a Lyapunov spectrum for $E=5$.
The initial values of $\theta_{1}(t)$, $\theta_{2}(t)$, $p_{\theta_{1}}(t)$, and $p_{\theta_{2}}(t)$ are 
taken as $\theta_{1}(0)=\theta_{2}(0)=\pi/2$, $p_{\theta_{1}}(0)=3$, and $p_{\theta_{2}}(0)=\sqrt{10}$\,.
Then the maximum Lyapunov exponent is $0.1999\pm0.0075$\,. 

%%%%%%%%%%%%%%%%%%%%%%%

\subsubsection*{(iii) $k=10$ case}

\begin{figure}[htbp]
\vspace*{0.7cm}
\centering
\begin{tabular}{c}
\begin{minipage}{0.45\hsize}
\centering
\includegraphics[width=7cm]{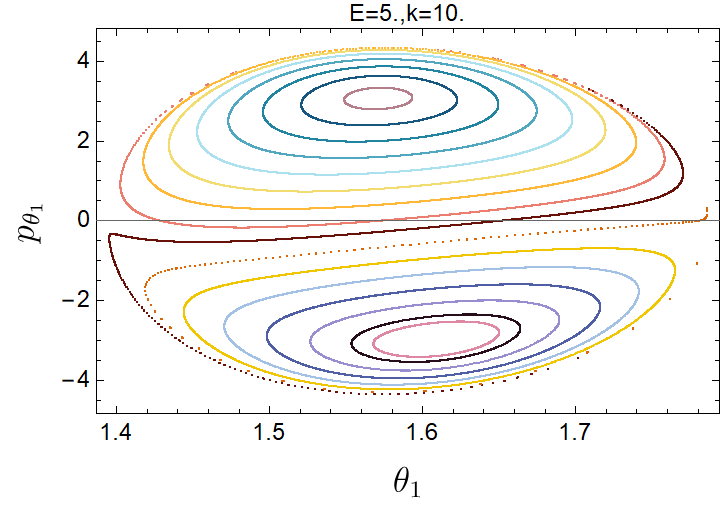}
\subcaption{\footnotesize Poincar\'{e} sections for $E=5$}
\label{fig:k10e5}
\end{minipage}
\begin{minipage}{0.45\hsize}
\centering
\includegraphics[width=7cm]{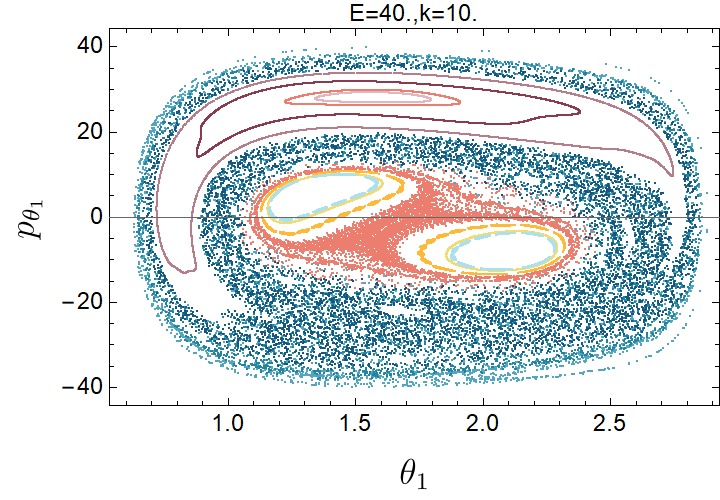}
\subcaption{\footnotesize Poincar\'{e} sections for $E=40$}
\label{fig:k10e40}
\end{minipage}
\vspace*{0.7cm}\\
\begin{minipage}[b]{0.45\hsize}
\centering
\includegraphics[width=7cm]{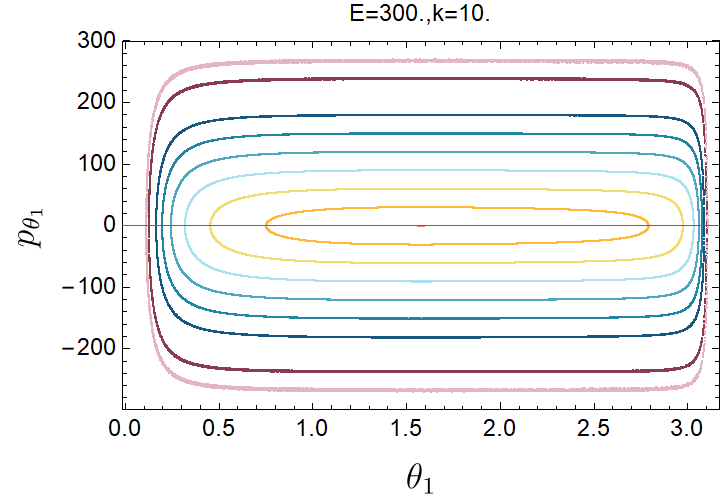}
\subcaption{\footnotesize Poincar\'{e} sections for $E=300$}
\label{fig:k10e300}
\end{minipage}
\begin{minipage}[b]{0.45\hsize}
\centering
\includegraphics[width=7cm]{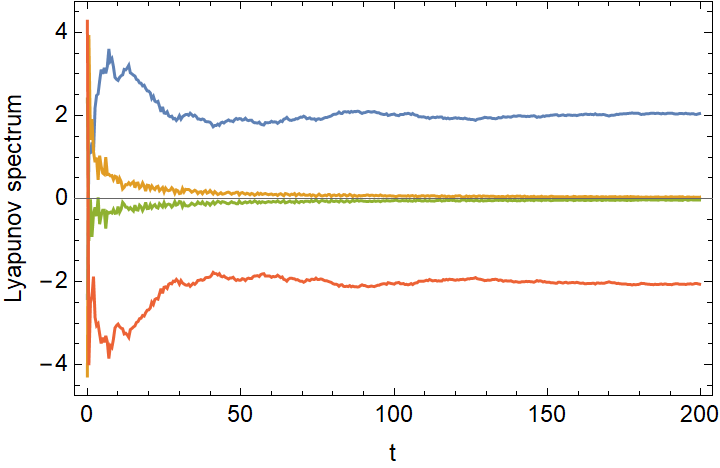}
\subcaption{\footnotesize Lyapunov spectrum for $E=40$, $p_1=15$}
\label{fig:lyapuk10}
\end{minipage}
\end{tabular}
\vspace*{0.7cm}
\caption{Poincar\'{e} sections and Lyapunov spectrum for $k=10$}
\label{fig:k=10}
\end{figure}

Figure \ref{fig:k=10} shows Poincar\'{e} sections for $k=10$\,. 
The potential minimum is $(\theta_{1},\theta_{2})\sim(\pi/2,\, \pi/2)$\,.
When $E=5$\,, there are only KAM tori and no chaos is found [Fig.\,\ref{fig:k=10} (a)]\,.
When $E=40$\,, some of KAM tori collide with each other and chaos appears [Fig.\,\ref{fig:k=10} (b)]\,.
As the energy becomes much higher $(E=300)$\,, there are only KAM tori and chaos disappears again [Fig.\,\ref{fig:k=10} (c)]\,. 
Notice here that higher energy is necessary to see chaos in this case in comparison 
to the previous two cases. 
This is simply because the large value of $k$ means a strong magnetic flux. 
The magnetic flux acts to force the particles to move in a circle and hence the KAM tori can survive even at relatively high energy. 

\medskip

Figure \ref{fig:k=10} (d) is a Lyapunov spectrum with $E=40$\,. 
The initial values of $\theta_{1}(t)$, $\theta_{2}(t)$\,, $p_{\theta_{1}}(t)$\,, 
and $p_{\theta_{2}}(t)$ are taken as $\theta_{1}(0)=\theta_{2}(0)=\pi/2$\,, $p_{\theta_{1}}(0)=15$\,, 
and $p_{\theta_{2}}(0)=37$\,. Then the maximum Lyapunov exponent is $2.0341\pm0.0182$\,.

%%%%%%%%%%%%%%%%%%%%%%%

\clearpage
%%%%%%%%%%%%%%%%
\section{Conclusion and Discussion}
\label{sec:conclusions}
%%%%%%%%%%%%%%%%

In this letter, we have studied a string sigma model on $R\times \mathcal{T}^{(\lambda_1,\lambda_2,\lambda)}_{k}$ by using a winding string ansatz. We have shown the presence of classical chaos for $k \neq \lambda^2$ by looking at Poincar\'{e} sections. We have also calculated the Lyapunov spectra 
and showed that the largest Lyapunov exponent is positive for chaotic string motions. 
Thus we conclude that the string sigma model is not classically integrable 
when $k \neq \lambda^2$\ as conjectured in \cite{Arutyunov:2020sdo}.

\medskip 

In our analysis, we have supposed a supergravity embedding of the background (\ref{eq:T11B})\,. 
It is a significant future problem to find out this supergravity embedding for general values of the parameters. 
Notably, in \cite{Arutyunov:2020sdo}, the background (\ref{eq:T11B}) has been derived as a certain limit 
of a more general background. It is a challenging problem to try to find out a supergravity embedding 
of it. It is also a nice practice to study non-integrable parameter region of it by following the standard manner 
as performed in this letter.

%%%%%%%%%%%%%%%%%%%%%%%%%%%
\section*{Acknowledgments}
\noindent
%%%%%%%%%%%%%%%%

We would like to thank H.~Y.~Chen for useful discussions during the online workshop 
on ``Online 2020 NTU-Kyoto high energy physics workshop.''  
We also thank Osamu Fukushima for fruitful discussions.
The work of T.I.\ was supported by JSPS Grant-in-Aid for Scientific Research (B) No.\,JP18H01214 and (C) No.\,JP19K03871.
The work of K.Y.\ was supported by the Supporting Program for Interaction-based Initiative Team Studies (SPIRITS) from Kyoto University, and JSPS Grant-in-Aid for Scientific Research (B) No.\,18H01214. This work is also supported in part by the JSPS Japan-Russia Research Cooperative Program.

%%%%%%%%%%%
%% References
%%%%%%%%%%%


\begin{thebibliography}{99}

\bibitem{Maldacena:1997re}
J.~M.~Maldacena,
``The Large N limit of superconformal field theories and supergravity,''
Int.\ J.\ Theor.\ Phys.\ \textbf{38} (1999) 1113-1133
%doi:10.1023/A:1026654312961
[arXiv:hep-th/9711200 [hep-th]].

\bibitem{Witten:1998qj}
E.~Witten,
``Anti-de Sitter space and holography,''
Adv.\ Theor.\ Math.\ Phys.\ \textbf{2} (1998) 253-291
%doi:10.4310/ATMP.1998.v2.n2.a2
[arXiv:hep-th/9802150 [hep-th]].

\bibitem{Gubser:1998bc}
S.~S.~Gubser, I.~R.~Klebanov and A.~M.~Polyakov,
``Gauge theory correlators from noncritical string theory,''
Phys.\ Lett.\ B \textbf{428} (1998) 105-114
%doi:10.1016/S0370-2693(98)00377-3
[arXiv:hep-th/9802109 [hep-th]].

\bibitem{Page:1984ae}
D.~N.~Page and C.~N.~Pope,
``Which Compactifications of $D=11$ Supergravity Are Stable?,''
Phys.\ Lett.\ B \textbf{144} (1984) 346-350
%doi:10.1016/0370-2693(84)91275-9

\bibitem{Romans:1984an}
L.~J.~Romans,
``New Compactifications of Chiral $N=2 d=10$ Supergravity,''
Phys.\ Lett.\ B \textbf{153} (1985) 392-396
%doi:10.1016/0370-2693(85)90479-4

\bibitem{Candelas:1989js}
P.~Candelas and X.~C.~de la Ossa,
``Comments on Conifolds,''
Nucl.\ Phys.\ B \textbf{342} (1990) 246-268
%doi:10.1016/0550-3213(90)90577-Z

\bibitem{Klebanov:1998hh}
I.~R.~Klebanov and E.~Witten,
``Superconformal field theory on three-branes at a Calabi-Yau singularity,''
Nucl.\ Phys.\ B \textbf{536} (1998) 199-218
%doi:10.1016/S0550-3213(98)00654-3
[arXiv:hep-th/9807080 [hep-th]].

\bibitem{Basu:2011di}
P.~Basu and L.~A.~Pando Zayas,
``Chaos rules out integrability of strings on AdS$_5 \times T^{1,1}$,''
Phys.\ Lett.\ B \textbf{700} (2011) 243-248
%doi:10.1016/j.physletb.2011.04.063
[arXiv:1103.4107 [hep-th]].

\bibitem{Basu:2011fw}
P.~Basu and L.~A.~Pando Zayas,
``Analytic Non-integrability in String Theory,''
Phys.\ Rev.\ D \textbf{84} (2011) 046006
%doi:10.1103/PhysRevD.84.046006
[arXiv:1105.2540 [hep-th]].

\bibitem{Asano:2015qwa}
Y.~Asano, D.~Kawai, H.~Kyono and K.~Yoshida,
``Chaotic strings in a near Penrose limit of AdS$_{5} \times$ T$^{1,1}$,''
JHEP \textbf{08} (2015) 060
%doi:10.1007/JHEP08(2015)060
[arXiv:1505.07583 [hep-th]].


\bibitem{Crichigno:2014ipa}
P.~M.~Crichigno, T.~Matsumoto and K.~Yoshida,
``Deformations of $T^{1,1}$ as Yang-Baxter sigma models,''
JHEP \textbf{12} (2014), 085
%doi:10.1007/JHEP12(2014)085
[arXiv:1406.2249 [hep-th]].

\bibitem{Klimcik1}
  C.~Klimcik,
  ``Yang-Baxter sigma models and dS/AdS T duality,''
  JHEP {\bf 0212} (2002) 051
  %doi:10.1088/1126-6708/2002/12/051
  [hep-th/0210095].

\bibitem{Klimcik2}
  C.~Klimcik,
  ``On integrability of the Yang-Baxter sigma-model,''
  J.\ Math.\ Phys.\  {\bf 50} (2009) 043508
  %doi:10.1063/1.3116242
  [arXiv:0802.3518 [hep-th]].

\bibitem{DMV1}
  F.~Delduc, M.~Magro and B.~Vicedo,
  ``On classical $q$-deformations of integrable sigma-models,''
  JHEP {\bf 1311} (2013) 192
  %doi:10.1007/JHEP11(2013)192
  [arXiv:1308.3581 [hep-th]].


\bibitem{DMV2}
  F.~Delduc, M.~Magro and B.~Vicedo,
  ``An integrable deformation of the AdS$_5\times$S$^5$ superstring action,''
  Phys.\ Rev.\ Lett.\  {\bf 112} (2014) no.5,  051601
  %doi:10.1103/PhysRevLett.112.051601
  [arXiv:1309.5850 [hep-th]].

\bibitem{KMY1}
  I.~Kawaguchi, T.~Matsumoto and K.~Yoshida,
  ``Jordanian deformations of the AdS$_5\times$S$^5$ superstring,''
  JHEP {\bf 1404} (2014) 153
  %doi:10.1007/JHEP04(2014)153
  [arXiv:1401.4855 [hep-th]].

\bibitem{MY}
  T.~Matsumoto and K.~Yoshida,
 ``Yang-Baxter sigma models based on the CYBE,''
  Nucl.\ Phys.\ B {\bf 893} (2015) 287
 % doi:10.1016/j.nuclphysb.2015.02.009
  [arXiv:1501.03665 [hep-th]]. 

\bibitem{SY-T11}
J.~Sakamoto and K.~Yoshida,
``Yang-Baxter deformations of $W_{2,4}\times T^{1,1}$ and the associated T-dual models,''
Nucl. Phys. B \textbf{921} (2017), 805-828
%doi:10.1016/j.nuclphysb.2017.06.017
[arXiv:1612.08615 [hep-th]]. 

\bibitem{Rado:2020yhf}
L.~Rado, V.~O.~Rivelles and R.~S\'anchez,
``Yang-Baxter deformations of the $ADS_5$ x $T^{1,1}$ superstring and their backgrounds,''
JHEP \textbf{02} (2021), 126
%doi:10.1007/JHEP02(2021)126
[arXiv:2010.14081 [hep-th]].

\bibitem{Crichigno:2015sga}
P.~Marcos Crichigno, T.~Matsumoto and K.~Yoshida,
``Towards the gravity/CYBE correspondence beyond integrability -- Yang-Baxter deformations of $T^{1,1}$,''
J. Phys. Conf. Ser. \textbf{670} (2016) no.1, 012019
%doi:10.1088/1742-6596/670/1/012019
[arXiv:1510.00835 [hep-th]].

\bibitem{Arutyunov:2020sdo}
G.~Arutyunov, C.~Bassi and S.~Lacroix,
``New integrable coset sigma models,''
JHEP \textbf{03} (2021), 062
%doi:10.1007/JHEP03(2021)062
[arXiv:2010.05573 [hep-th]].


\bibitem{Guadagnini:1987ty}
E.~Guadagnini, M.~Martellini and M.~Mintchev,
``SCALE  INVARIANCE SIGMA MODELS ON HOMOGENEOUS SPACES,''
Phys.\ Lett.\ B \textbf{194} (1987) 69
%doi:10.1016/0370-2693(87)90771-4

\bibitem{PandoZayas:2000he}
L.~A.~Pando Zayas and A.~A.~Tseytlin,
``Conformal sigma models for a class of $T^{(p,q)}$ spaces,''
Class.\ Quant.\ Grav.\ \textbf{17} (2000) 5125-5131
%doi:10.1088/0264-9381/17/24/312
[arXiv:hep-th/0007086 [hep-th]]. 

\bibitem{Levine:2021fof}
N.~Levine and A.~A.~Tseytlin,
``Integrability vs. RG flow in $G \times G$ and $G \times G /H$ sigma models,''
[arXiv:2103.10513 [hep-th]].

\bibitem{Shimada:1979}
I.~Shimada and T.~Nagashima,
``A Numerical Approach to Ergodic Problem of
Dissipative Dynamical Systems,''
 Prog.\ Theor.\ Phys.\ \textbf{61} (1979) 1605.

\bibitem{Kolmogorov:1954}
A.~N.~Kolmogorov,
``The conservation of conditionally periodic motion with a small variation in the Hamiltonian,''
Dokl.\ Akad.\ Nauk SSSR \textbf{98} (1954) 527.

\bibitem{Arnold:1963}
V.~I.~Arnold,
``Small denominators and problems of stability of motion in classical and celestial mechanics,''
Uspekhi Mat.\ Nauk, Russian Math.\ \textbf{18} No.\ 6 (1963) 91; Russ.\ Math.\ Surv.\ 
\textbf{18} (1963) 9.

\bibitem{Moser:1962}
J.~Moser,
``On invariant curves of area-preserving mappings of an annulus,''
 Nachr.\ Akad.\ Wiss.\ G$\ddot{\rm o}$ttingen Math.-Phys.\ Kl.\ II (1962) 1.





\end{thebibliography}
\end{document}